\documentclass[aps,prl,twocolumn,superscriptaddress,floatfix,nofootinbib,longbibliography]{revtex4-2}
\usepackage{epsfig,amsmath,amssymb,color,comment,physics}
\usepackage[normalem]{ulem}
\usepackage[english]{babel}
\usepackage{dsfont}
\usepackage{float}
\usepackage[bookmarks=true,colorlinks,linkcolor=RoyalBlue,urlcolor=RoyalBlue,citecolor=RoyalBlue]{hyperref}
\usepackage[dvipsnames]{xcolor}
\usepackage{soul} 
\usepackage{color, xcolor}  

\hypersetup{
    unicode=false,     % non-Latin characters in Acrobat bookmarks
    pdftoolbar=false,  % show Acrobat toolbar?
    pdfmenubar=true,   % show Acrobatmenu?
    pdffitwindow=false, % window fit to page when opened
    pdfstartview={FitH},% fits the width of the page to the window
    pdftitle={},    % title
    pdfauthor={Authors},     % author
    pdfsubject={},   % subject of the document
    pdfcreator={},   % creator of the document
    pdfproducer={}, % producer of the document
    pdfnewwindow=true,% links in new window
    colorlinks=true,% false: boxed links; true: colored links
    linkcolor=black,% color of internal links (change box color with linkbordercolor)
    citecolor=blue, % color of links to bibliography
    filecolor=magenta,% color of file links
    urlcolor=blue% color of external links
}

\newcommand{\beginsupplement}{%
    \setcounter{table}{0}
    \renewcommand{\thetable}{S\arabic{table}}%
    \setcounter{figure}{0}
    \renewcommand{\thefigure}{S\arabic{figure}}%
    \setcounter{equation}{0}
    \renewcommand{\theequation}{S\arabic{equation}}%
    \setcounter{section}{0}
    \renewcommand{\thesection}{S\arabic{section}}%
   }

\begin{document}

\title{Quantum Mpemba effect without global symmetries}

\author{Tanmay Bhore}
\affiliation{School of Physics and Astronomy, University of Leeds, Leeds LS2 9JT, United Kingdom}

\author{Lei Su}
\affiliation{Department of Physics, University of Chicago, Chicago, Illinois 60637, USA}

\author{Ivar Martin}
\affiliation{Materials Science Division, Argonne National Laboratory, Lemont, Illinois 60439, USA}
\affiliation{Department of Physics, University of Chicago, Chicago, Illinois 60637, USA}

\author{Aashish A. Clerk}
\affiliation{Pritzker School of Molecular Engineering, University of Chicago, Chicago, Illinois 60637, USA}

\author{Zlatko Papi\'c}
\affiliation{School of Physics and Astronomy, University of Leeds, Leeds LS2 9JT, United Kingdom}

\date{\today}
\begin{abstract}
The Mpemba effect, where a system initially farther from equilibrium relaxes faster than one closer to equilibrium, has been extensively studied in classical systems and recently explored in quantum settings. While previous studies of the quantum Mpemba effect (QME) have largely focused on isolated systems with global symmetries, we argue that the QME is ubiquitous in generic, non-integrable many-body systems lacking such symmetries, including U(1) charge conservation, spatial symmetries, and even energy conservation. Using paradigmatic models such as the quantum Ising model with transverse and longitudinal fields, we show that the QME can be understood through the energy density of initial states and their inverse participation ratio in the energy eigenbasis. Our findings provide a unified framework for the QME, linking it with classical thermal relaxation.

\end{abstract}
\maketitle

{\bf \em Introduction.---}The approach to thermal equilibrium is the ultimate fate of most interacting systems in nature, both classical and quantum. The rate at which this happens, however, can vary wildly depending on the initial microscopic configurations and the nature of interactions. A celebrated example, dating back to antiquity and systematically studied since the 1960s, is the classical Mpemba effect~\cite{Mpemba1969}: the counterintuitive observation that a hot liquid can freeze faster than a colder liquid under certain conditions. 
While similar anomalous relaxation behaviors have been found in a variety of classical systems~\cite{Keller18,BaityJesi2019,Kumar2020,Schwarzendahl2022,GonzalesAdalid2024}, the understanding of their origin remains an active area of research~\cite{BurridgeLinden2016, Klich2019,Bechhoefer2021, Holtzman2022, Teza2025}. 

In contrast, quantum counterparts of the Mpemba effect have only recently garnered attention~\cite{Ares2025Review}. Much like its classical analog, the quantum Mpemba effect (QME) can be defined as an anomalous dynamical process where a system, initially further from equilibrium, relaxes faster than a system that starts closer to equilibrium. In open quantum systems, the QME can be engineered by suppressing the slowest eigenmode of the Liouvillean~\cite{Carollo2021,Zhang2025}, reminiscent of classical Markovian systems~\cite{Klich2019,Kumar2020}. A variety of other mechanisms have also been recently explored~\cite{Kochsiek2022,bao2022acceleratingrelaxationmarkovianopen,Ivander2023,Zhou2023,Chatterjee2023,Chatterjee2024,Wang2024,Liu2024,Longhi2024,wang2024mpembameetsquantumchaos,furtado2024strongquantummpembaeffect,qian2024intrinsicquantummpembaeffect,Bettmann2025,Dong2025,Nava2024,medina2024anomalousdischargingquantumbatteries,kheirandish2024mpembaeffectquantumoscillating,graf2025roleelectronelectroninteractionmpemba,Zatsaryna2025,strachan2024nonmarkovianquantummpembaeffect,wang2024goingquantummarkovianityreality, westhoff2025fastdirectpreparationgenuine}. Quantum simulations of the QME in the presence of Markovian dynamics have been performed with trapped ions~\cite{Shapira2024,Zhang2025} and superconducting qubits~\cite{edo2024studyquantumthermalizationthermal}.

In this work, we instead focus on \emph{closed} quantum systems evolving under unitary dynamics. For such systems, the QME can be conveniently probed by a global quantum quench: the system is prepared in a pure state -- one that is not an eigenstate of the Hamiltonian -- and then evolved under the Schr\"odinger equation. In this setting, the process of equilibration is understood at the level of subsystems approaching the Gibbs ensemble at late times~\cite{DeutschETH,SrednickiETH,RigolNature}. Signatures of the QME have been found in many examples of closed quantum systems, e.g., various types of spin models~\cite{Ares2023_Mpemba_NatureComms,Murciano2024XY, Rylands2024Heisenberg}, free fermions and bosons~\cite{Yamashika2024,yamashika2024quenchingsuperfluidfreebosons}, quantum circuits~\cite{turkeshi2024quantummpembaeffectrandom,klobas2024translationsymmetryrestorationrandom,Liu_Mpemba_PRL_2024,yu2025symmetrybreakingdynamicsquantum,ares2025entanglementasymmetrydynamicsrandom,Klobas2024Rule54,Foligno2025}, many-body localized systems~\cite{liu2024quantummpembaeffectsmanybody}, and trapped ion experiments~\cite{Joshi_Mpemba_PRL}. In particular, for integrable models, Ref.~\cite{Rylands_Mpemba_PRL_2024} developed a comprehensive understanding of the QME based on the quasiparticle picture of the spreading of correlations following the quench~\cite{CalabreseCardy2005}. Nevertheless, these studies have mainly focused on a restricted class of models possessing a global U(1) symmetry, such as the conservation of particle number or spin magnetization. This raises a fundamental question: is the QME possible in the absence of global symmetries?

In this paper, we give an affirmative answer to the previous question by demonstrating that the QME can occur in generic many-body systems that 
 are non-integrable and lack global symmetries, such as U(1) charge, spatial symmetries, or even energy conservation.  We consider the paradigmatic quantum Ising model in the presence of transverse and longitudinal fields, in both its Hamiltonian and Floquet incarnations, as well as local spin-chain models with random couplings. In all cases, we show that the QME can be succinctly explained by taking into account (i) the energy density of the initial states -- playing the role of effective temperature, and (ii) the inverse participation ratio of the initial states in the energy eigenbasis. Our results point to the ubiquity of the QME in generic chaotic systems and provide a unified description that highlights the role of energy distributions.

{\bf \em Mixed-field Ising model.---}As our first example, we consider the one-dimensional (1D) mixed-field Ising model,
\begin{equation}\label{eq:MFIM}
    H^\mathrm{MFIM} = J_{zz} \sum_{i = 1}^{N-1} {\sigma}^z_i {\sigma}^z_{i+1} + h_x \sum_{i = 1}^{N} {\sigma}^x_i + h_z \sum_{i=2}^{N-1} {\sigma}^z_i,
\end{equation}
where $\sigma_j^{x,z}$ denote standard Pauli matrices on $j$th site and $h_x$, $h_z$ are the transverse and longitudinal fields, respectively. We assume an open chain and include boundary fields, $\delta h_1^z = 0.25$ and $\delta h_N^z = -0.25$, to break reflection symmetry. The Hamiltonian (\ref{eq:MFIM}) then lacks any symmetries except energy conservation. We fix the coupling strengths to $(J_{zz}, h_z, h_x) = (1, (1 + \sqrt{5})/4, (\sqrt{5} + 5)/8)$, for which the model is chaotic and shows ballistic growth of entanglement and diffusive energy transport~\cite{Kim2013}, although we expect the results described here to hold for generic parameter choices for which the model is chaotic.

To reveal the QME, we consider translation-invariant product states:
\begin{equation}
    |\theta,\phi\rangle = \bigotimes_{i=1}^N \Big[ \cos(\theta/2) \ket{\uparrow}_i + e^{i \phi} \sin(\theta/2) \ket{\downarrow}_i \Big],
    \label{eq:states}
    \end{equation}
where each spin points at the same angle $(\theta,\phi)$ on the Bloch sphere. 
In isolated systems energy is conserved, hence we can assign a temperature to a state $\ket{\Psi}$ via its energy, $E_{\Psi} = \langle{\Psi}|{H}|\Psi\rangle = \frac{1}{\mathcal{Z}}\mathrm{Tr}(H e^{-\beta {H}})$, where $\beta = 1/T$ is the inverse temperature in units $k_B = 1$ and ${\mathcal{Z}}=\mathrm{Tr}(e^{-\beta {H}})$. Infinite temperature or $\beta=0$ corresponds to a uniform average of the eigenenergies, while energies away from the middle of the spectrum correspond to $\beta\neq 0$, i.e., finite (either positive or negative) temperatures. 

\begin{figure}
    \includegraphics[width=\linewidth]{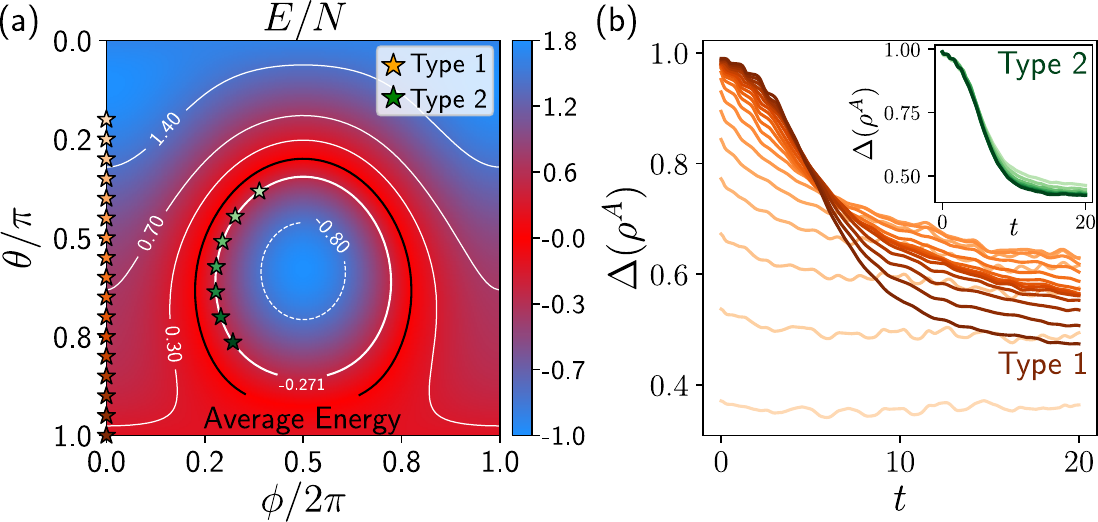}
    \caption{QME in the mixed field Ising model, Eq.~(\ref{eq:MFIM}). (a): Energy density of  tilted ferromagnetic states in Eq.~(\ref{eq:states}) as a function of angles $\theta$, $\phi$, with isothermal levels marked by the contours. States with increasing energy (Type 1) and states at the same temperature (Type 2) are marked by the orange and green stars, respectively. As a guide to the eye, we also indicated the average energy (black curve). (b): Dynamics of trace distance, Eq.~(\ref{eq:trace_distance}), for the subsystem density matrix $\rho^A$ following the quench from Type 1 initial states, with Type 2 states shown in the inset. The lines correspond to the star symbols in panel (a). The QME is visible for states of Type 1 but it is absent for states of Type 2. Results are for system size $N = 14$ obtained by exact diagonalization. The subsystem $A$ is chosen to be one half of the chain. 
    }
    \label{fig:Ising_energy_dynamics}
\end{figure}
The energy density of $\ket{\theta,\phi}$ states admits a simple form~\cite{SM}, which we plot as a function of $(\theta,\phi)$ in Fig.~\ref{fig:Ising_energy_dynamics}(a). The color scale marks the effective temperature of the states; energies close to the edges of the spectrum are depicted in blue color, corresponding to low (positive or negative) temperatures, while the temperature progressively rises towards the middle of the spectrum, with the average energy corresponding to infinite temperature (red color). Among the states in Eq.~(\ref{eq:states}), we consider two special types. Type 1 denotes  ferromagnetic states with $\phi = 0$ that are rotated around the $y$-axis, which progressively decreases their energy density. States of Type 2 are initial states degenerate in energy and away from the middle of the spectrum, thus at a fixed finite temperature.  

Next, in Fig.~\ref{fig:Ising_energy_dynamics}(b) we quench the system from Type 1 and 2 initial states and monitor the evolution of trace distance~\cite{Shapira2024,Zhang2025} as a diagnostic of the QME: 
\begin{equation}\label{eq:trace_distance}
    \Delta(\rho^A)(t) = \frac{1}{2}||\rho^A(t) - \rho_\mathrm{DE}(\rho^A)||_1.
\end{equation}
Here, $\rho^A = \mathrm{Tr}_B[\rho(t)]$ denotes the reduced density matrix for the bipartition of the system into two halves $A$ and $B$, $\rho_\mathrm{DE}(\rho^A) = \sum_E \bra{E} \rho(0) \ket{E} \mathrm{Tr}_B(\ket{E}\bra{E})$ is the diagonal ensemble~\cite{dAlessio2016} for the initial state $\rho(0)$, restricted to $A$ via partial trace, and $\ket{E}$ are the energy eigenstates. The partial trace is required in Eq.~(\ref{eq:trace_distance}) as the full state of the system -- assumed to be pure initially -- cannot evolve to a mixed state under unitary dynamics otherwise.  Note that many previous studies used a different diagnostic of the QME -- the ``entanglement asymmetry''~\cite{Ares2023_Mpemba_NatureComms,russotto2025symmetrybreakingchaoticmanybody,Ares2025Review}, however we employ a more general metric that does not rely on the presence of any global symmetry. In the Supplementary Material (SM)~\cite{SM} we show that $\Delta(\rho^A)$ is a good diagnostic of the QME  for U(1)-conserving systems, along with other measures, such as the Frobenius distance~\cite{Ares2025Review}, which yield qualitatively similar conclusions.  

For small tilt angles $ \theta$, Type 1 states are close to the edge of the spectrum and show almost no dynamics in Fig.~\ref{fig:Ising_energy_dynamics}(b). Upon increasing $\theta$, the initial trace distance progressively increases, along with its rate of decay. This eventually leads to a crossing between successive states around $t\approx 4$ in Fig.~\ref{fig:Ising_energy_dynamics}(b). This is a manifestation of the QME, wherein hotter states relax to equilibrium faster than colder states. 
Note that in a chaotic system,  $\Delta(\rho^A)(t\to\infty) $ is expected to approach $0$ for the initial state at a finite energy density as the system size $N \to \infty$. The deviation seen in Fig.~\ref{fig:Ising_energy_dynamics}(b) is attributed to a finite-size effect, we perform a system-size scaling in the Supplementary Material (SM)~\cite{SM}. At the same time, we observe that the states of Type 2, which are degenerate in average energy, show little variation in their trace distance dynamics, with essentially identical initial trace distance and decay rates [inset of Fig.~\ref{fig:Ising_energy_dynamics}(b)]. 
This is consistent with the classical intuition where the Mpemba effect requires two initial states at different temperatures.

{\bf \em The role of energy distributions.---}In integrable systems, the QME arises due to states that are simultaneously more out-of-equilibrium and have larger overlaps with the fastest quasiparticle modes~\cite{Rylands_Mpemba_PRL_2024}. Similar mechanisms have been used to explain the effect in systems coupled to external reservoirs~\cite{Ares2025FreeFermionsMixed,Caceffo2024}. However, our model (\ref{eq:MFIM}) is chaotic and lacks a quasiparticle description at high energies. In the absence of such a description, we instead look at the eigenstates themselves to understand the conditions for the QME. 

Consider our initial pure state $|\Psi\rangle\equiv |\theta,\phi\rangle$. In the energy eigenbasis, the time-evolved density matrix can be decomposed into off-diagonal and diagonal components:
\begin{equation}\label{Eq:density_matrix}
    |\Psi(t) \rangle \langle \Psi(t)| = \rho_\mathrm{DE} + \sum_{n \neq m} c_{n} c_m^* e^{-i(E_n - E_m)t} | E_n \rangle \langle E_m|,
\end{equation}
where $\rho_\mathrm{DE}$ is the diagonal ensemble and $c_n = \langle E_n | \Psi \rangle$ encodes the energy overlaps of the initial state. Partial tracing both sides to restrict to a subsystem, we get
\begin{equation}\label{Eq:trace_dist_off_diagonal}
\Delta(\rho^A)(t) = \frac{1}{2}\Big|\Big|  \sum_{n \neq m} c_{n} c_{m}^* e^{-i(E_n - E_m)t}\mathrm{Tr_B} (| E_n \rangle \langle E_m|)  \Big|\Big|_1.
\end{equation}
The QME requires two conditions to be met: two states should have unequal initial trace distances, and the state with the larger initial trace distance must relax to thermal equilibrium faster than the other state. By Eq.~\eqref{Eq:trace_dist_off_diagonal}, the trace distance at $t =0$ is equal to $1/2||\sum_{m \neq n}c_n c_m^* \mathrm{Tr_B}(|E_n \rangle \langle E_m|) ||_1$, which is a sum over $\mathcal{O}(\mathcal{D}^2)$ eigenstate pairs, each weighed by the off-diagonal energy overlaps, where $\mathcal{D}$ denotes the Hilbert space dimension. If the initial state is an eigenstate, $\Delta(\rho^A)$ remains constant at all times and dynamics is frozen. If the initial state involves  more eigenstates, a larger number of overlaps $c_n$ contribute to the sum, which can lead to a larger initial trace distance. At the same time, a larger energy spread implies that a wider range of frequencies $E_n - E_m$ enter the exponent in Eq.(\ref{Eq:trace_dist_off_diagonal}), leading to faster dephasing and eventual thermalization. We expect this dephasing process to be the fastest at infinite temperature, wherein an $\mathcal{O}(\mathcal{D}^2)$ frequencies enter the exponent. 

\begin{figure}[tb]
    \includegraphics[width=\linewidth]{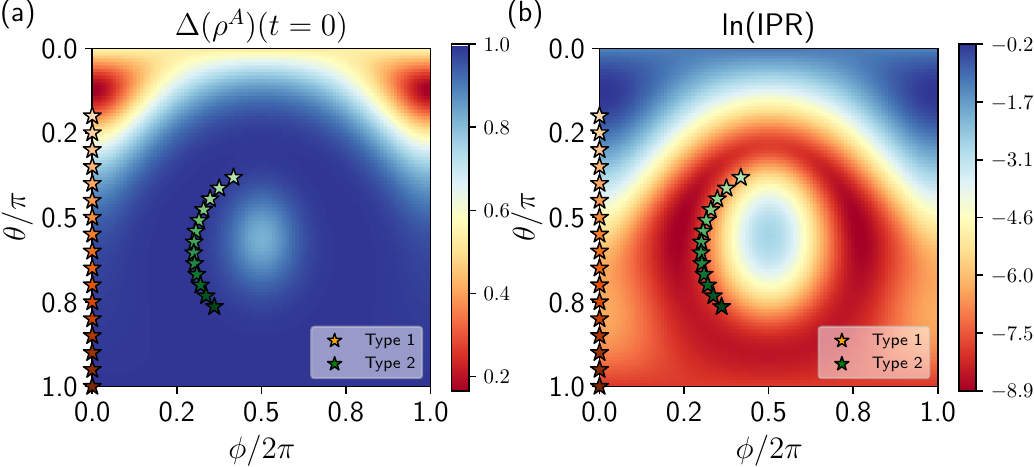}
    \caption{Origin of the QME in the mixed field Ising model, Eq.~(\ref{eq:MFIM}). (a) Initial trace distance from the diagonal ensemble and (b) Logarithm of the inverse participation ratio (IPR) of $\ket{(\theta, \phi)}$ states, Eq.~(\ref{eq:states}). IPR anti-correlates with the initial trace distance: states with smaller IPR have larger initial trace distances and vice versa. Results are for $N = 14$.}
    \label{fig:IPR_tracedist}
\end{figure}

Delocalization in the energy eigenbasis can then correlate with both conditions required for the QME. A simple quantity that quantifies this delocalization 
is the inverse participation ratio (IPR)~\cite{Thouless1974}, 
\begin{equation}\label{eq:IPR}
 \mathrm{IPR}(\ket{\Psi}) = \sum_E |\langle E|\Psi \rangle|^4,  
\end{equation}
where we assumed the state $\ket{\Psi}$ is normalized. The IPR takes values between $[1/\mathcal{D}, 1]$, where the limits correspond to a fully delocalized state (infinite temperature) and a single energy eigenstate, respectively. Since smaller IPR values correspond to more delocalized states, we expect an anti-correlation between IPR values and the initial trace distance, along with the speed of its decay.

We plot the initial trace distance $\Delta(\rho^A)(t=0)$ and the logarithm of IPR for $\ket{\theta, \phi}$ states in Fig.~\ref{fig:IPR_tracedist}(a)-(b). We observe a clear anti-correlation between the two, with regions corresponding to low IPR values showing large initial trace distances and vice versa.  In particular, this explains the trace distance dynamics observed in Fig.~\ref{fig:Ising_energy_dynamics}. The initial $\Delta(\rho^A)$ for Type 1 states increases while the IPR monotonically decreases with increasing $\theta$, leading to the QME. By contrast, Type 2 states, which are degenerate in energy, have almost constant IPR values; this reflects in the almost-similar trace distance dynamics and the absence of the QME in the inset of Fig.~\ref{fig:Ising_energy_dynamics}(b). 

{\bf \em QME in random Hamiltonians.---}To reveal the significance of energy distributions in the QME, we now construct a random model that allows us to tune the energy density and the IPR of an ensemble of initial states. We consider states in Eq.~(\ref{eq:states}) with random angles on each site, i.e., $\theta_i$, $\phi_i$ are independently drawn from the uniform distribution over a fraction $f$ of its full domain. For a small $f$, the ensemble is close to a polarized state at a fixed energy, whereas $f=1$ corresponds to an infinite temperature ensemble.

If the spins roughly all point in the same direction, any local term of the form $(\boldsymbol{\sigma}_i \times \boldsymbol{\sigma}_{i+1})\cdot\mathbf{\hat v}$ has a zero expectation value, where $\mathbf{\hat{v}}$ is the unit vector along the three spin directions. In fact, the spectrum of this term is centered around zero, making all polarized states infinite temperature states for such Hamiltonians. To tune their energy density, we need to add terms that assign them a constant energy, e.g., an isotropic Heisenberg term of the form $\boldsymbol{\sigma}_i\cdot \boldsymbol{\sigma}_{i+1}$. Finally, a desired Hamiltonian is:
\begin{equation}\label{eq:random_ham}
    {H}_\mathrm{R} = \sum_{i=1}^{N-1} (\boldsymbol{\sigma}_i \times \boldsymbol{\sigma}_{i+1})\cdot \hat{\mathbf{v}}_i + J_\mathrm{H} \sum_{i}^{N-1} \boldsymbol{\sigma}_i\cdot\boldsymbol{\sigma}_{i+1},
\end{equation}
where the three components $\mathbf{v}_i = (v_x,v_y,v_z)_i$ are real numbers drawn uniformly from the window $[-1,1]$ and we set $J_\mathrm{H}=-4$. The random vectors $\hat{\mathbf{v}}_i$ ensure that the model breaks all global symmetries and it is non-integrable, as confirmed by its average level spacing ratio matching the Wigner-Dyson prediction~\cite{OganesyanHuse2007}. 

\begin{figure}[tb]
    \includegraphics[width=0.5\textwidth]{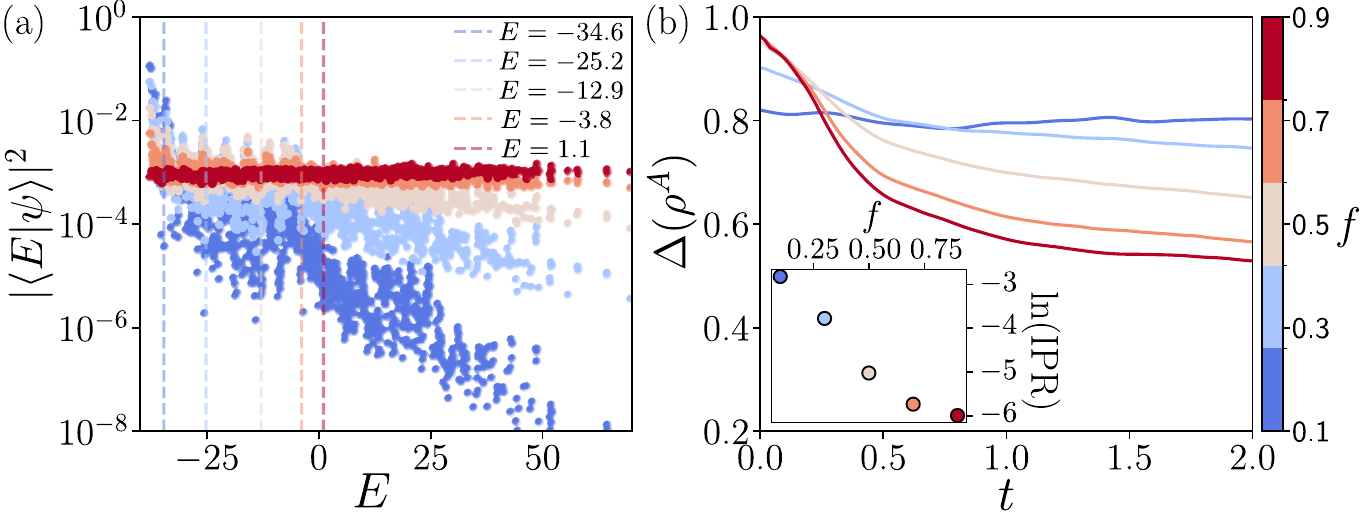}
    \caption{QME in the random model, Eq.~(\ref{eq:random_ham}). (a) Overlap of $\ket{(\theta_i,\phi_i)}$ states with energy eigenstates as a function of their energy. The overlaps are averaged over 100 realizations of the angles. Different colors represent different values of $f$ (see text), resulting in average energies given in the legend. (b) Trace distance dynamics for $\ket{(\theta_i,\phi_i)}$ states. The crossing associated with the QME is visible as the effective temperature of the state is tuned. The inset shows the IPR for the same states as a function of $f$. All data is for system size $N=10$ and fixed disorder window $W = 1$. 
    }
    \label{fig:random_hamiltonian_overlaps}
\end{figure}
In Fig.~\ref{fig:random_hamiltonian_overlaps}(a), we show the overlaps of random states $\ket{(\theta_i,\phi_i)}$ with energy eigenstates of $H_\mathrm{R}$, averaged over 100 realizations of the angles. For small $f$, the overlap peaks near the ground state, while in the limit $f\to1$, the overlap is evenly distributed across the spectrum, moving the state toward infinite temperature. 
This coincides with a flattening of the distribution, as reflected in the decreasing \text{IPR} shown in the inset of Fig.~\ref{fig:random_hamiltonian_overlaps}(b), marking delocalization over a larger number of eigenstates. Indeed, the QME is evident in the trace distance dynamics in Fig.~\ref{fig:random_hamiltonian_overlaps}(b), where increasing $f$ leads simultaneously to higher initial $\Delta(\rho^A)$ values and faster decay rates, resulting in a crossing.

{\bf \em Floquet systems.---}Up to this point, we have focused on Hamiltonian systems with energy conservation. However, our previous considerations based on the IPR can be directly generalized to the Floquet eigenstates of a periodically-driven system. To demonstrate the QME in the absence of energy conservation, we consider a Floquet version of Eq.~(\ref{eq:MFIM}) -- the kicked Ising  model, with the Floquet operator given by~\cite{Prosen1998,ProsenJPA1998,Bertini2019entanglement}
\begin{eqnarray}\label{eq:floquet_MFIM}
\nonumber U_\mathrm{KI} &=& \exp\Big[-iT_1 ( J_{zz} \sum_{j} {\sigma}^z_j {\sigma}^z_{j+1} +  h_z \sum_{j} {\sigma}^z_j )\Big] \\
&& \times \exp\Big[ -i T_2 \; h_x \sum_{j}{\sigma}^x_j \Big],   
\end{eqnarray}
where we set the periods $T_1=T_2=1/2$ for simplicity. 

For $h_z \neq 0$, $U_\mathrm{KI}$ describes a chaotic model. Due to ``Floquet heating"~\cite{Bukov2015},  
for a generic initial condition, 
the state of the subsystem $A$ then approaches the infinite-temperature thermal ensemble at late times, $\sigma^A =  \mathds{1}/\mathcal{D}_A$, where $\mathcal{D}_A$ is the Hilbert space dimension of $A$. This simplification in the Floquet case allows us to consider the relative entropy $S(\rho^A(t)||\sigma^A)$~\cite{Ares2025Review}
as a diagnostic of the QME, as the latter reduces to the entanglement entropy, $S_E =-\mathrm{Tr}\rho^A(t)\ln\rho^A(t)$, up to an unimportant constant factor involving $\mathcal{D}_A$.

For any pure and unentangled initial state, the initial relative entropy from $\sigma^A$ is the same. Hence, to observe the QME in the Floquet case, we go beyond product states and consider weakly entangled initial states in the class of translation-invariant infinite matrix product states (iMPS) of the form $    \ket{\Psi(A)} = \sum_{\mathbf{s}}(...A^{[s_{i-1}]} A^{[s_i]}A^{[s_{i+1}]}...)\ket{\mathbf{s}}$, where $\ket{\mathbf{s}}$ denotes the basis states~\cite{CiracRMP}. For our purposes, it is sufficient to consider a single-parameter familty of $2\times 2$ MPS matrices $A$:
\begin{align}\label{eq:MPS}
A^{[\downarrow]}(\theta) = 
\begin{pmatrix}
\cos\theta & 0 \\
\sin\theta & 0
\end{pmatrix}, \quad
A^{[\uparrow]}(\theta) = 
\begin{pmatrix}
0 & -i \\
0 & 0
\end{pmatrix}.
\end{align}
This parametrization yields a normalized iMPS and it has been used to describe quench dynamics in a constrained Ising model arising in Rydberg atom arrays~\cite{Ho_PRL_2019,Daniel2023}.

Since entanglement growth is unbounded in an infinite Floquet system, the QME dynamics can be achieved if states with higher initial entanglement show a slower rate of entropy growth than states with smaller initial entanglement. 
To test this, we time evolve the states in Eq.~(\ref{eq:MPS}) using the infinite time-evolving block decimation algorithm (iTEBD) \cite{Vidal_PRL_2007} and evaluate the entanglement entropy dynamics for various values of $\theta$ in Fig.(\ref{fig:Floquet_QME})(a). With increasing $\theta$, the initial  entanglement in the state grows while its rate of growth simultaneously decreases, leading to multiple crossings in the $S_E$ curves. This exemplifies the QME dynamics through the lens of entanglement entropy. To test the role of quasi-energy distributions, we truncate the initial states to a finite chain of length $N$, and calculate their IPR in the Floquet eigenbasis in Fig.(\ref{fig:Floquet_QME})(b). The IPR behavior indeed correlates with the crossings in entropy similar to previous results, e.g., Fig.~\ref{fig:random_hamiltonian_overlaps}(b) of the random model.

\begin{figure}[tb]
\includegraphics[width=0.98\linewidth]{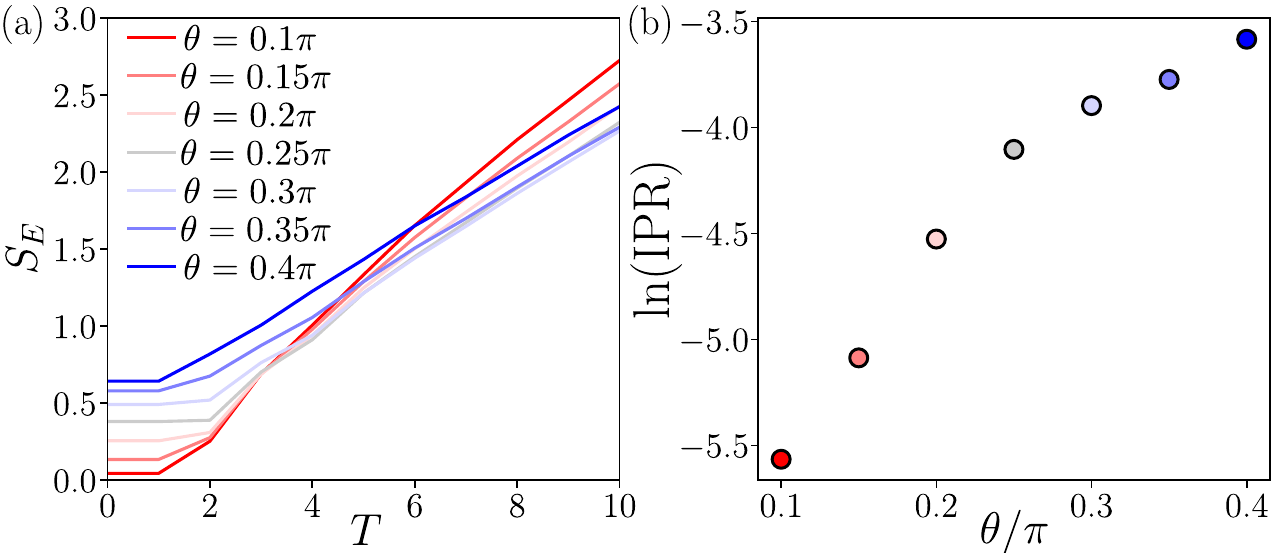}
    \caption{QME in the kicked Ising model, Eq.~(\ref{eq:floquet_MFIM}). (a) Entanglement entropy dynamics of iMPS states (\ref{eq:MPS}) in integer multiples of the Floquet period $T=T_1+T_2=1$ for several values of $\theta$. The QME occurs as $\theta$ is increased. Data is obtained using iTEBD for the same parameters as specified below Eq.~(\ref{eq:MFIM}) and the bond dimension is $\chi=128$. (b) Logarithm of IPR for a finite version of MPS states against the Floquet eigenstates obtained via exact diagonalization for the system size $N=10$ with open boundary conditions. 
    }\label{fig:Floquet_QME} 
\end{figure}

{\bf \em Conclusions.---}We demonstrated the ubiquity of the QME in a wide class of isolated quantum systems without special symmetries hitherto assumed. We elucidated a common origin of the QME based on energy distributions of initial states. This not only reproduces the earlier results for systems with global symmetries, but it also allows to construct the QME in systems without such symmetries, including chaotic and random models. In the SM~\cite{SM}, we further demonstrate the generality of our conclusions by considering different choices of norms, initial states, and other measures. In particular, our findings shed new light on previous observations of ``strong'' vs. ``weak thermalization''~\cite{Banuls2011,LinMotrunichIsing2017} in the same model (\ref{eq:MFIM}), which now can be interpreted as an instance of the QME.  

One interesting open question is: what is the universal diagnostic of the QME? In the SM~\cite{SM}, we confirm that simple spectral measures, such as energy variance and higher moments, generally do not account for our observations~\cite{SM}. On the other hand, while the IPR is a good diagnostic for $\ket{\theta, \phi}$ product states as well as MPS states (\ref{eq:MPS}) considered above, the IPR diagnostic may break down for more highly-entangled states. As a counterexample, it is enough to consider 
two initially low-entangled states that  undergo time evolution {\em after}  trace distance crossing. During subsequent evolution, they will no longer exhibit the QME; yet, their IPR remains the same as it was at $t=0$. We thus conjecture that, as a diagnostic of the QME, the IPR is limited to weakly entangled initial states. Formulating a precise general criterion remains an interesting open question.    

\begin{acknowledgments}

{\bf Acknowledgments.---} We thank Andrew Hallam, Matthew Yusuf, Sara Murciano and, in particular, Jie Ren for useful discussions. Some of the numerical simulations were performed using \href{https://github.com/jayren3996/iTEBD.jl}{iTEBD.jl} package. T.B. and Z.P. acknowledge support by the Leverhulme Trust Research Leadership Award RL-2019-015 and EPSRC Grant EP/Z533634/1. 
Work of L.S. and I.M., particularly on Floquet QME,  was supported by the US Department of Energy, Office of
Science, Basic Energy Sciences, Materials Sciences and Engineering Division.
A.C. acknowledges support from the Simons Foundation (Grant No. 669487).
This research was supported in part by grant NSF PHY-2309135 to the Kavli Institute for Theoretical Physics (KITP). The data used in this paper is publicly available at~\cite{data_manuscript}.
\end{acknowledgments}

\bibliography{refs.bib}

\newpage 
\cleardoublepage 

\beginsupplement

\newpage

\setcounter{equation}{0}
\setcounter{figure}{0}
\setcounter{table}{0}
\setcounter{page}{1}
\setcounter{section}{0}
\makeatletter
\renewcommand{\theequation}{S\arabic{equation}}
\renewcommand{\thefigure}{S\arabic{figure}}
\renewcommand{\thesection}{S\Roman{section}}
\renewcommand{\thepage}{\arabic{page}}
\renewcommand{\thetable}{S\arabic{table}}

\onecolumngrid

\begin{center}
\textbf{\large Supplemental Online Material for ``Quantum Mpemba effect without global symmetries" }\\[5pt]
\vspace{0.1cm}
\begin{quote}
{\small In this Supplementary Material, we study several diagnostics of the quantum Mpemba effect (QME) and demonstrate that our choice of measure in the main text reproduces the previous results obtained using entanglement asymmetry in models with conserved U(1) charge. Furthermore, we support our conclusions in the main text for different types of initial states, including weakly-entangled states without translation invariance. Finally, we demonstrate that previous findings of ``weak'' and ``strong'' thermalization in the mixed field Ising model~\cite{Banuls2011} can also be described in the framework of the QME. 
}  \\[20pt]
\end{quote}
\end{center}

\maketitle

\onecolumngrid

\section{Entanglement asymmetry and trace distance}

Most previous studies of the quantum Mpemba effect (QME) assumed models with a global symmetry, such as  U(1), and the quantity dubbed ``entanglement asymmetry'' was introduced as a witness of the  QME~\cite{Ares2023_Mpemba_NatureComms,Ares2025Review}. Intuitively, entanglement asymmetry quantifies the restoration of symmetry in the reduced matrix describing a subsystem of the full system when the latter is quenched from an initial state that breaks the symmetry.  

For a Hamiltonian with a global symmetry (assumed to be Abelian), consider a charge operator $Q$ which may be decomposed as $Q = Q_A + Q_B$ for a given bipartitioning of the system. For any initial state $\rho$ which commutes with $Q$, the reduction of $\rho$ onto a subsystem by partial trace produces a density matrix that is block-diagonal in the basis of $Q_A$. Consider an initial state $\rho$ which breaks the symmetry such that $[\rho, Q] \neq 0$. Then, at the level of the subsystem, $[\rho^A,Q_A] \neq 0$, implying that $\rho^A$ is not a block-diagonal matrix in the basis of $Q_A$, but generally contains off-block-diagonal elements. The entanglement asymmetry is then defined as
\begin{equation}
    \Delta S_A \equiv  S(\rho^{A,Q}) - S(\rho^A), \quad S(\rho) = -\mathrm{Tr}(\rho \ln \rho), \quad \rho^{A,Q} = \sum_q \mathcal{P}_q \rho^A\mathcal{P}_q,
\end{equation}
where $\mathcal{P}_q$ is a projector onto the eigenspace of $Q_A$ with charge $q$. If $[\rho,Q] = 0$, then $\rho^A$ is block-diagonal in the eigenbasis of $Q_A$ and $S(\rho^{A,Q}) = S(\rho^A)$, thus the entanglement asymmetry vanishes, whereas $\Delta S_A >0$ for a state that breaks the symmetry. Note that entanglement asymmetry can be written as a relative entropy: $\Delta S_A = S(\rho^A|| \rho^{A,Q}) \equiv \mathrm{Tr} [\rho^A (\ln \rho^A - \ln \rho^{A,Q})]$. 

The degree to which a state breaks the symmetry can be quantified by the norm of the commutator $[\rho, Q]$. Generally, we expect that if a state breaks the symmetry more, it will have a larger value of $\Delta S_A$, making it a \emph{bona fide} measure of symmetry breaking.
In 1D short-range interacting systems, the Hohenberg-Mermin-Wagner theorem states that continuous symmetries, such as U(1), cannot be spontaneously broken at any finite temperature. This implies that a symmetry, which is broken initially, should  be restored in thermal equilibrium, if thermal equilibrium is reached dynamically. Note that there are some known exceptions in systems with non-Abelian symmetries~\cite{ares2023lack}.  Furthermore, integrable or many-body localized systems do not thermalize. However, for ``generic'', non-disordered short-range interacting systems, we expect entanglement asymmetry to behave similarly to other  measures of distance from thermal equilibrium~\cite{Ares2025Review}. 

\begin{figure}[bth]
    \includegraphics[width=\textwidth]{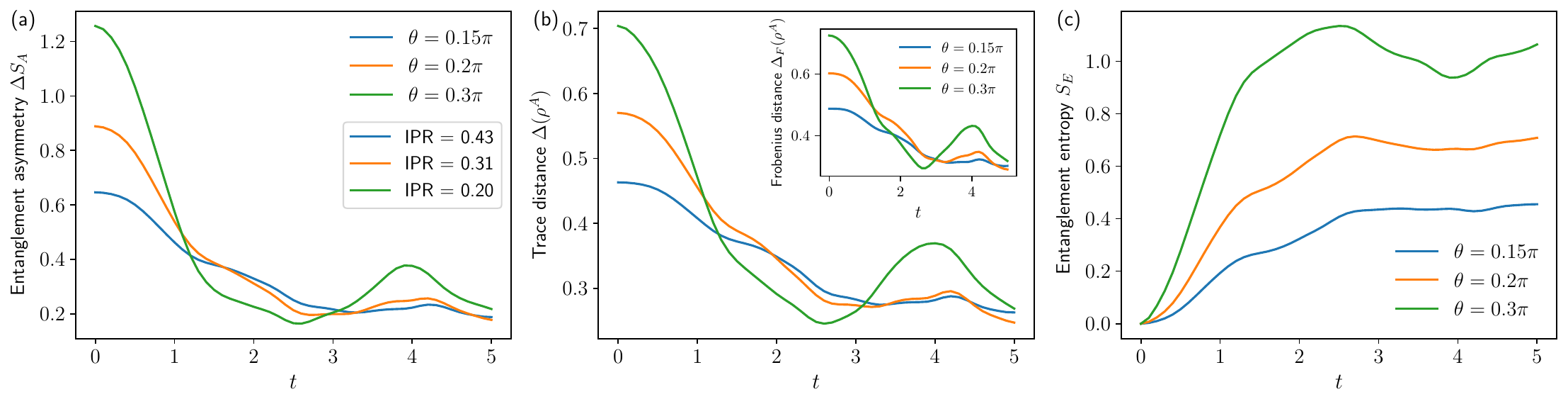}
    \caption{(a) Dynamics of entanglement asymmetry, (b) trace distance and Frobenius distance (inset) and (c) entanglement entropy for the chaotic XXZ model in Eq.~(\ref{suppeq:XXZ}) and ferromagnetic states in Eq.~(\ref{suppeq:tiltedFM}) tilted along the $y$-axis. Results are obtained by exact diagonalization for a system size $N = 10$ with $J_1 = J_2 = 1$ and $\Delta_1 = \Delta_2 = 0.5$.}
    \label{fig_sup:Mpemba_EA_comaprison}
\end{figure}
In Fig.~\ref{fig_sup:Mpemba_EA_comaprison} we compare the trace distance of the initial reduced density matrix $\Delta(\rho_A)$, as defined in the main text, to the entanglement asymmetry $ \Delta S_A$ in a chaotic short-ranged model with U(1) symmetry. 
We consider the XXZ Hamiltonian with next-nearest-neighbor couplings that break integrability, previously studied in Ref.~\cite{Ares2023_Mpemba_NatureComms}:
\begin{equation}\label{suppeq:XXZ}
        H = J_1 \sum_{i=1}^{N-1} \left({\sigma}^x_i {\sigma}^x_{i+1} + {\sigma}^y_i {\sigma}^y_{i+1} + \Delta_1 {\sigma}^z_i {\sigma}^z_{i+1}\right) + J_2 \sum_{i=1}^{N-2} \left({\sigma}^x_i {\sigma}^x_{i+2} + {\sigma}^y_i {\sigma}^y_{i+2} + \Delta_2 {\sigma}^z_i {\sigma}^z_{i+2}\right).
\end{equation}
This Hamiltonian has a U(1) symmetry as it conserves $M=\sum_i \sigma_i^z$; the initial breaking and restoration of this symmetry is measured by the entanglement asymmetry. Specifically, we consider tilted ferromagnetic states: 
\begin{equation}\label{suppeq:tiltedFM}
 \ket{\psi(\alpha; \theta)} = \exp(-i \frac{\theta}{2}\sum_i \sigma_i^\alpha)\ket{\uparrow\uparrow\uparrow....}, \; \alpha = x,y,z,  
\end{equation}
which break the U(1) symmetry for $\alpha=x,y$ and non-zero $\theta$. We consider an equal bipartition of the system and calculate the entanglement asymmetry, trace and Frobenius distance (as defined in Eq. (\ref{eq:Frob})) and the entanglement entropy for three different values of $\theta$ in Fig.~\ref{fig_sup:Mpemba_EA_comaprison}. We observe that the trace and Frobenius distances qualitatively reproduce the behavior of the entanglement asymmetry and show the QME with increasing $\theta$. This shows that the trace and Frobenius distance metrics used in the main text successfully captures the QME in systems with symmetries, considered in previous work. 

We also find that the IPR of the states in Fig.~\ref{fig_sup:Mpemba_EA_comaprison} follows the expected trend as we observed in systems without symmetries in the main text, i.e., smaller IPR corresponding to a larger initial trace distance and simultaneously a faster rate of decay. Curiously, we also observe a difference in the slope of entanglement entropy, $S_E = -\mathrm{Tr}(\rho^A \ln \rho^A)$, for the same states. Although all three states have zero entanglement entropy initially, the states with slow trace distance dynamics also show slow growth and smaller saturation values of entanglement entropy. This also suggests a connection between the rate of entanglement entropy growth and the QME dynamics. We explore this connection further in the next section.

\section{Other diagnostics of QME}

\begin{figure}[tbh]
    \includegraphics[width=0.7\textwidth]{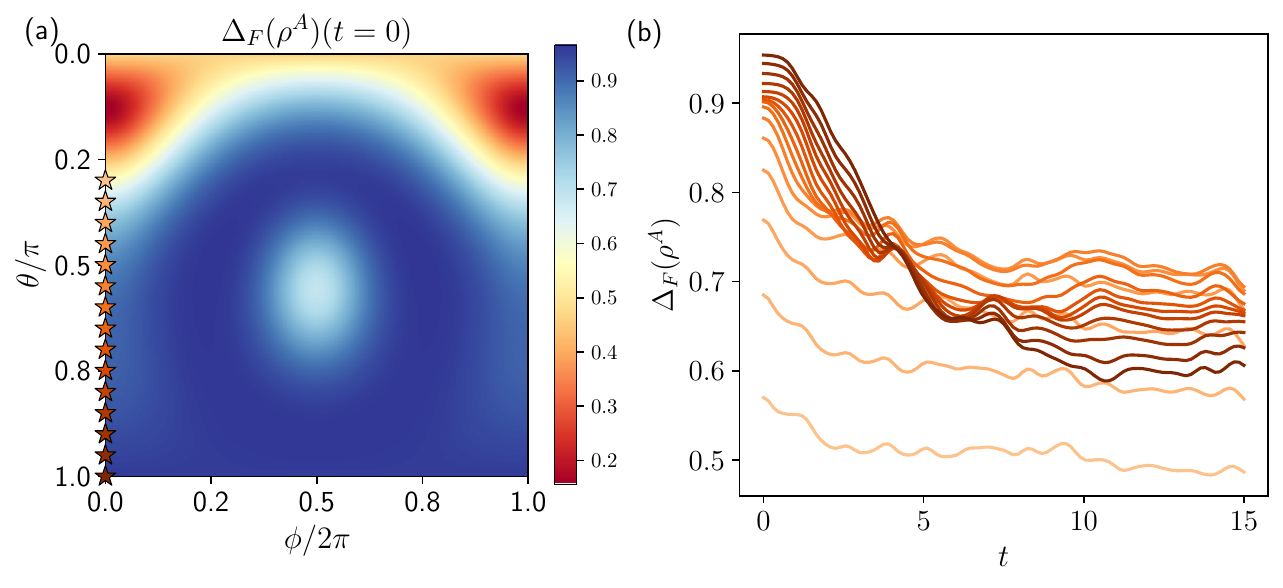}
    \caption{(a) Frobenius distance of tilted ferromagnetic states from the diagonal ensemble, reduced to a subsystem $A$ of $N/2$ sites in the mixed field Ising model. (b) Dynamics of the Frobenius distance for Type 1 initial states marked by stars in panel (a). Results are for system size $N = 10$ obtained by exact diagonalization.}
    \label{fig_sup:Mpemba_frob}
\end{figure}

Many possible metrics have been suggested to describe the QME dynamics, see Ref.~\cite{Ares2025Review} for a discussion. In the main text, we have employed the trace distance measure to compare the time-evolved density matrix with the diagonal ensemble. In contrast, recent experiments on trapped ions have characterized the QME dynamics using the Frobenius norm between the initial and the equilibrium density matrix~\cite{Joshi_Mpemba_PRL}. Here, we confirm that the Frobenius norm qualitatively reproduces the QME dynamics as observed with the trace distance metric in the main text.
The Frobenius distance of an initial state from the diagonal ensemble is defined as
\begin{align}\label{eq:Frob}
\Delta_F(\rho^A) = \sqrt{1 - \min\left(1, \; \frac{2 \, \mathrm{Tr}[\rho^A \rho_{\mathrm{DE}}^A]}{\mathrm{Tr}[(\rho^A)^2 + (\rho_{\mathrm{DE}}^A)^2]}\right)}.    
\end{align}
We consider $\ket{\theta, \phi}$ initial states defined in the main text and evaluate the initial Frobenius distance across the full range of $\theta$, $\phi$ values in Fig~\ref{fig_sup:Mpemba_frob}(a). The Frobenius distance metric behaves similarly to the trace distance metric in Fig.~2 of the main text. We also consider states of Type 1, marked by stars in Fig~\ref{fig_sup:Mpemba_frob}(a), and calculate the dynamics of their Frobenius distance in   Fig~\ref{fig_sup:Mpemba_frob}(b). Similar to the results in the main text, we find the QME dynamics in the Frobenius distance as $\theta$ is increased, pointing to its robustness against the choice of distance metrics.

\begin{figure}[tbh]
    \includegraphics[width=0.7\textwidth]{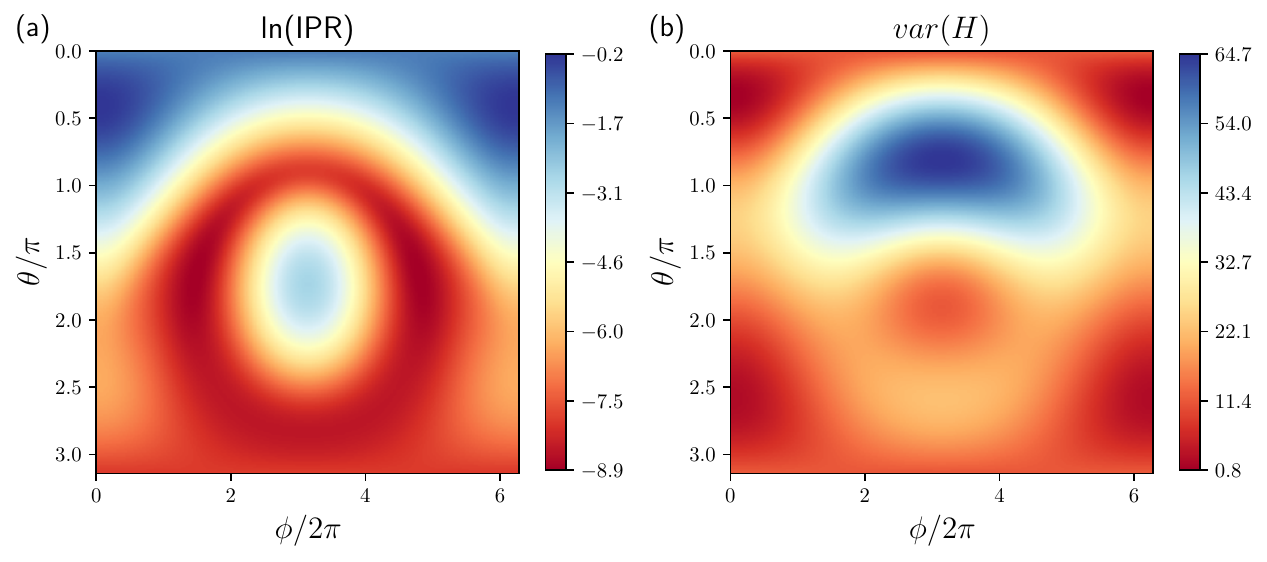}
    \caption{(a) IPR and (b) energy variance of the $\ket{\theta,\phi}$ states for the mixed field Ising Hamiltonian in the main text. 
    Unlike the IPR, the energy variance does not capture the states associated with the QME dynamics observed in Fig.~1 of the main text.  
    Results are obtained by exact diagonalization for the system size $N = 14$.}
    \label{fig_sup:IPR_variance}
\end{figure}

Given the crucial role of energy distributions on the QME dynamics, it is natural to ask if other measures of the energy distribution could act as reliable metrics of the QME. Here, we show that one of the simplest alternative measures -- the energy variance of initial states -- does not act as a good metric of the QME dynamics. 
First, the average energy of the $\ket{\theta,\phi}$ states in Eq.~(2) for the mixed field Ising Hamiltonian admits a simple form and is given by 
\begin{equation}
 \lim_{N\to\infty} \langle \hat{H}_\mathrm{MFIM}\rangle_{\ket{\theta,\phi}}/N = J_{zz} \cos^2(\theta) + h_x \sin(\theta) \cos(\phi) + h_z \cos(\theta).   
\end{equation}
We use the standard definition of the energy variance:
\begin{equation}
    var(H) \equiv \langle\hat{H}^2\rangle_{\Psi} - \langle \hat{H} \rangle^2_{\Psi}.
\end{equation}
In Fig. (\ref{fig_sup:IPR_variance})(a)-(b), we plot the IPR and the energy variance, $var(H)$, for the $\ket{\theta, \phi}$ states of Eq.~(2) of the main text for the mixed field Ising Hamiltonian in Eq.~(1). For the states of Type 1 and 2, considered in Fig.~1 of the main text, the IPR either decreases monotonically (Type 1) or stays constant (Type 2), whereas their energy variance shows no such correlation. This indicates that the energy variance does not act as a sensitive metric of the QME dynamics for the models considered in this paper.

Indeed, it can be argued that energy variance and IPR do not necessarily imply the same thing. The IPR quantifies delocalization over \textit{eigenstates},
and it is high (low) if the state is more localized (delocalized) over the energy eigenbasis. On the other hand, the energy variance, measuring how sharply peaked the energy distribution is around its mean, is high (low) if the energy distribution is broad (sharp). Naively, one might expect states with high energy variance to have a low IPR and vice versa.
To see why this need not always be the case, consider a state that is delocalized over an extensive number of eigenstates that all have nearly the same energy. Then, the IPR of the state is low, whereas the energy variance is small, as the energies are tightly clustered. Conversely, if the state is delocalized over two energy eigenstates with very different energies, the IPR is high but the energy variance is also large. Thus, the two quantities probe fundamentally different aspects of the energy distribution.

\begin{figure}
    \includegraphics[width=\textwidth]{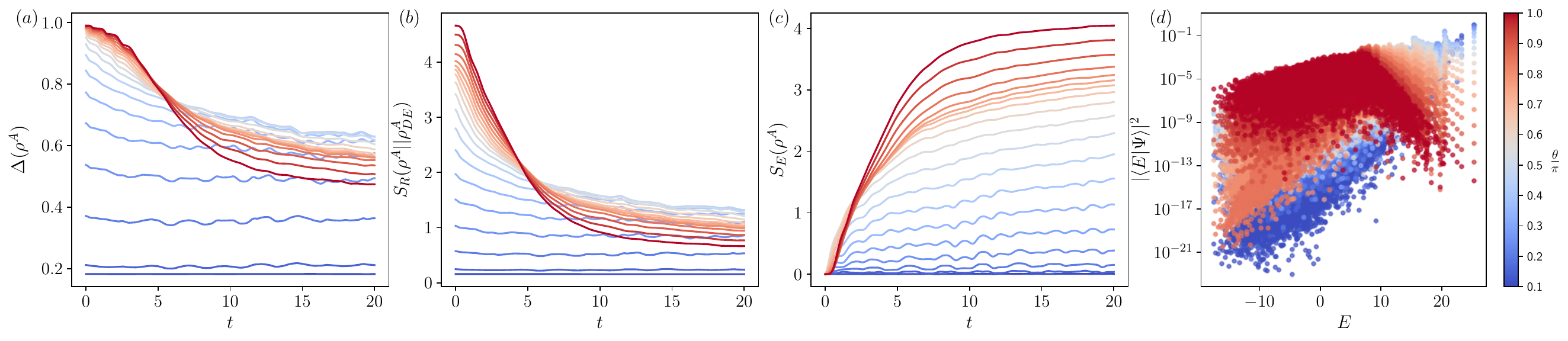}
    \caption{Dynamics of trace distance (a), relative entropy (b) and  entanglement entropy (c), with energy distributions of the initial states (d). All data is for the mixed field Ising Hamiltonian from the main text and tilted ferromagnetic states in Eq.~(\ref{suppeq:XXZ}). System size is $N = 14$ with a subsystem size of $7$ sites.}
    \label{fig_sup:Mpemba_rel_entropy_overlaps}
\end{figure}

While energy variance is insufficient to explain our results, we are able to draw connections with another standard measure of thermalization: the growth of entanglement entropy. Although infinite temperature states in chaotic quantum systems are expected to show ballistic growth of entanglement \cite{Kim2013}, the rate of growth can depend finely on the temperature of the initial states. Since the QME dynamics observed for states of Type 1 in the mixed field Ising model in the main text relies on varying the temperature of initial states, we expect to see this reflected in the rate of growth of entanglement entropy. 
We consider two entropic quantities; the entanglement entropy $S_E$ and the relative entropy of the state with its diagonal ensemble, $S_R(\rho^A|| \rho^A_{\mathrm{DE}}) = \mathrm{Tr} [\rho^A ( \ln \rho^A -\ln \rho_\mathrm{DE}^A)]$, where $\rho_\mathrm{DE}^A$ is the diagonal ensemble for $\rho$ reduced to subsystem $A$. 
We consider states in Eq.~(\ref{suppeq:tiltedFM}) tilted around the $y$-axis and evolve them under the mixed field Ising Hamiltonian in the main text. The trace distance $\Delta(\rho^A)$, relative entropy $S_R$ and entanglement entropy $S_E$ for varying $\theta$ are shown in Fig~\ref{fig_sup:Mpemba_rel_entropy_overlaps}(a)-(c). 
We observe that both the trace distance and the relative entropy capture the QME dynamics, with eventual crossings appearing as $\theta$ is increased. This growth is supplemented by the dynamics of entanglement entropy, where we observe an increasing rate of growth and saturation value of entropy as $\theta$ is increased. Increasing $\theta$ increases the effective temperature of the state and we expect the fastest growth and saturation to the Page value of entropy at infinite temperature~\cite{PhysRevLett.71.1291}. 

To gain a finer understanding of these results, we also look at the energy eigenstate overlaps of these initial states, shown in Fig~\ref{fig_sup:Mpemba_rel_entropy_overlaps}(d).
For small $\theta$, the initial state has the largest overlap on the most excited eigenstate. With increasing $\theta$, the weight of the initial state on the most excited state consistently decreases, while the spectrum simultaneously flattens. This delocalization is marked by the monotonic decrease in IPR values, as previously seen in  Fig.~2 of the main text.

\section{Finite-size scaling of trace distance dynamics}

\begin{figure}[tbh]
    \includegraphics[width=1\textwidth]{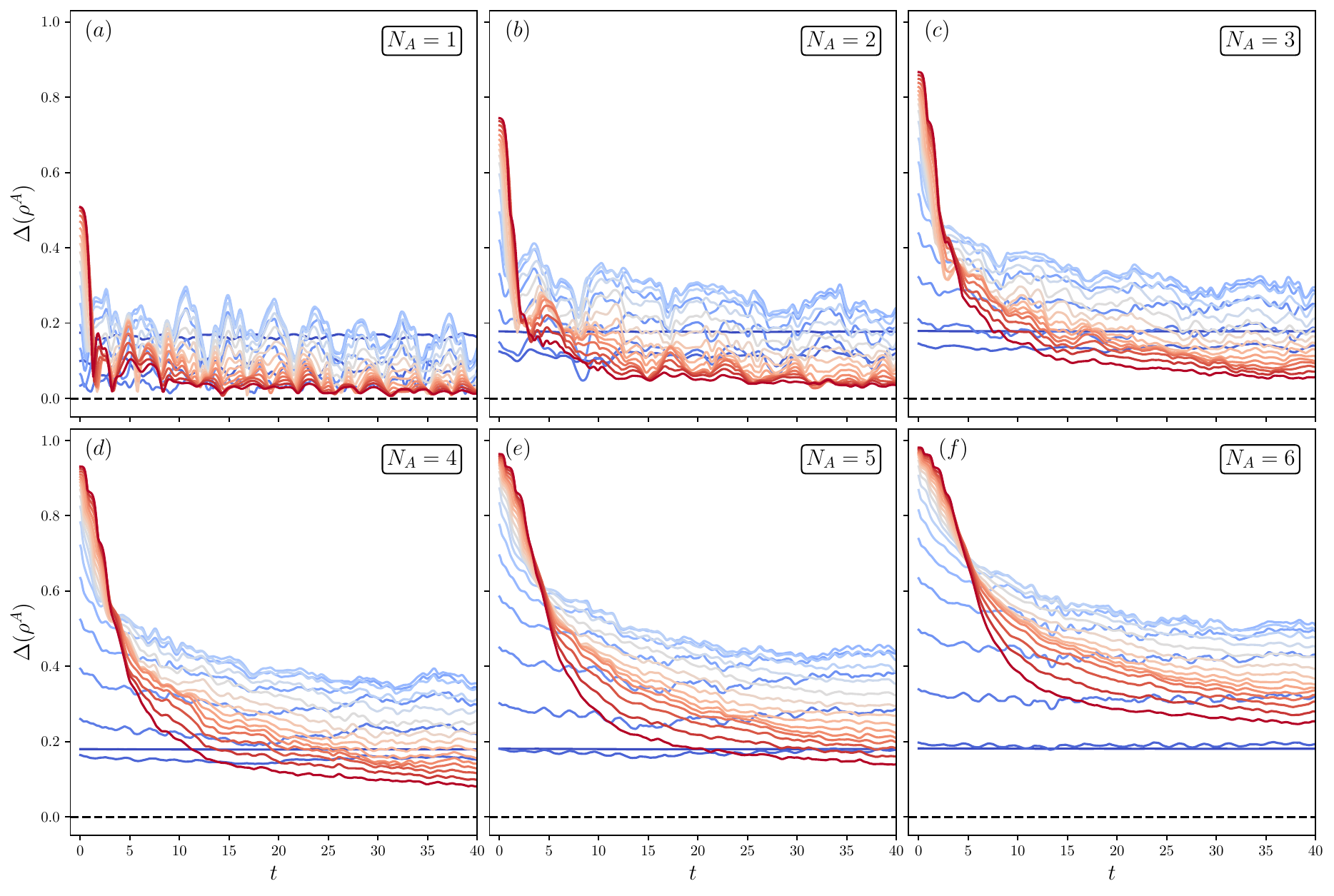}
    \caption{Trace distance dynamics of Type 1 initial states in the mixed field Ising model for increasing size of the subsystem $A$. The system size is $N = 14$ and the subsystem is chosen as the first $N_A$ spins from the left.}
    \label{fig_sup:trace_dist_scaling}
\end{figure} 

\begin{figure}[tbh]
    \includegraphics[width=0.5\textwidth]{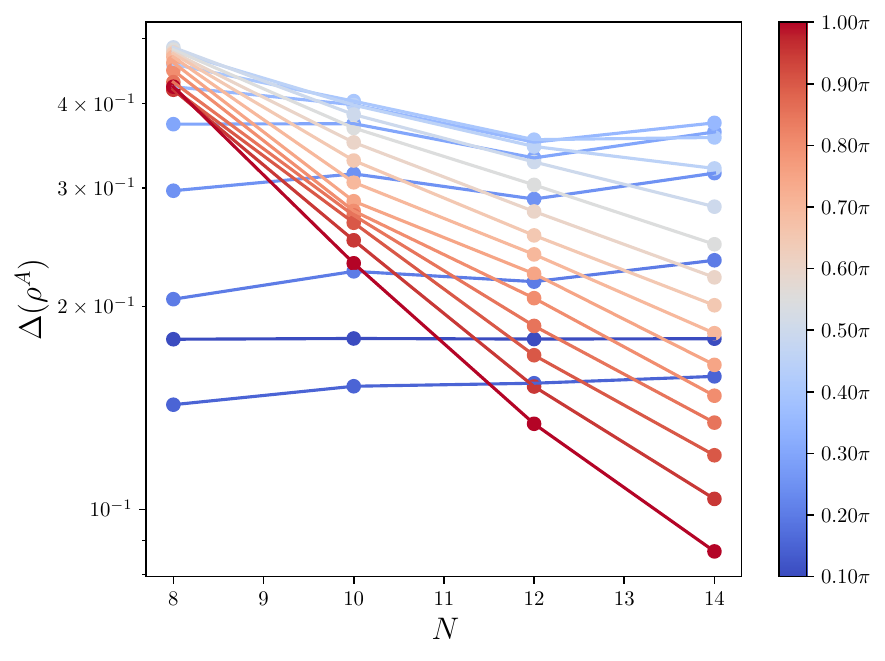}
    \caption{Saturation value of trace distance of Type 1 initial states in the mixed field Ising model with a fixed subsystem size $N_A = 4$ and increasing total system size $N$. While the saturation value plateaus with system size for small $\theta$, it drops to zero exponentially with $N$ for large $\theta$. 
    }
    \label{fig_sup:trace_dist_scaling_fixed_NA}
\end{figure} 

In this section, we perform a system-size scaling analysis of the trace distance dynamics for Type 1, i.e., tilted ferromagnetic initial states in the mixed field Ising model of Eq.~(1) in the main text. 
This scaling can be performed in two complementary ways, either by fixing the total system size $N$ and increasing the subsystem size $N_A$ up to the maximum of $N/2$, or by keeping $N_A$ fixed and scaling up $N$. The latter method gives us a way to access properties of a finite subsystem in the thermodynamic limit.
At first, we fix the total system size to $N = 14$ and consider the first $N_A$ spins from the left as our subsystem. The corresponding trace distance dynamics as $N_A$ is varied from 1 to 6 is shown in Fig. \ref{fig_sup:trace_dist_scaling}. 
Thermalization in isolated systems requires that the size of the bath be much bigger than the subsystem size for the subsystem to effectively thermalize. This is evident in Fig. \ref{fig_sup:trace_dist_scaling};
for small $N_A$, the effective bath is much larger than the subsystem and the trace distance $\Delta(\rho^A)$ drops close to zero for most initial states, whereas upon increasing the size of the subsystem, the trace distance plateaus further away from zero. However, QME is visible for all choices of $N_A$.

Next, we plot the scaling of saturation values of the trace distance for a fixed subsystem size, $N_A = 4$ as a function of $N$ in Fig.~\ref{fig_sup:trace_dist_scaling_fixed_NA}. For small $\theta$, the states remain close to the edge of the spectrum and the trace distance does not go to zero with increasing $N$, whereas it drops to zero exponentially with $N$ for large $\theta$ as the states delocalize over the energy eigenbasis.

\section{Other initial states}

\begin{figure}[tbh]
    \includegraphics[width=\textwidth]{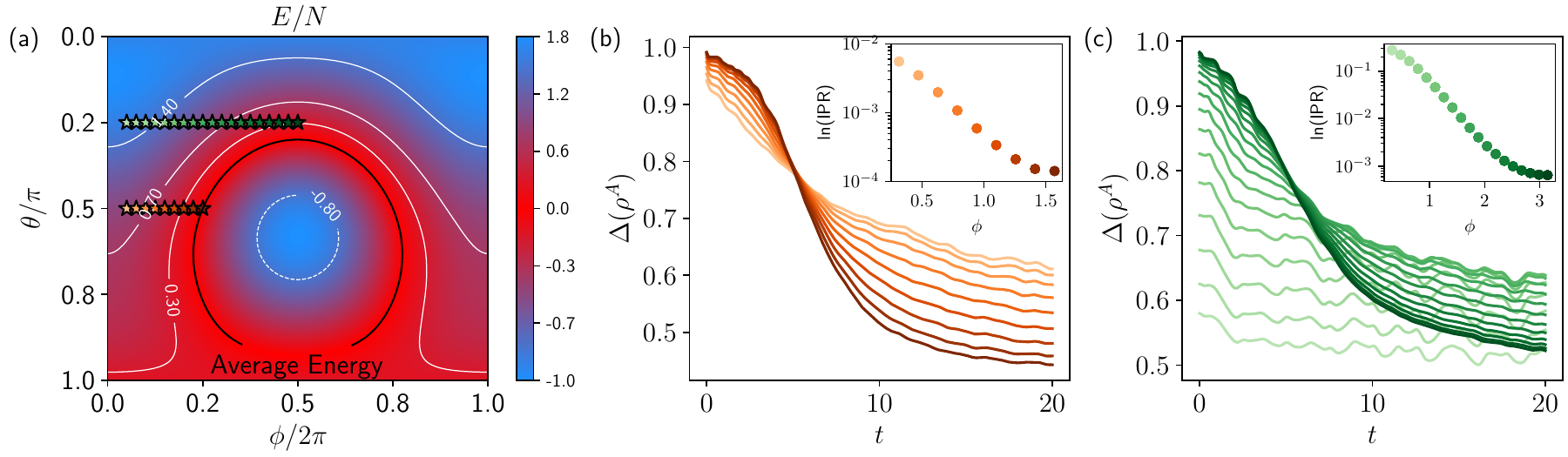}
    \caption{(a) Energy density of $\ket{\theta,\phi}$ states on the Bloch sphere along with two classes of states marked with orange and green stars. (b)-(c) Trace distance dynamics of the states marked by orange and green along with their IPRs(inset). The QME is visible in both the cases. Results are shown for $N = 14$.}
    \label{fig_sup:other_init}
\end{figure}

In the main text we have focused on two classes of translation-invariant product states: Type 1 states with increasing energies (effective temperature) and Type 2 states at a fixed energy or effective temperature. Given the correlation between IPR values and the initial tace distance in Fig.~2 of the main text, we expect to see the QME dynamics across a large range of $\theta,\phi$ values, so long as we stay away from the boundaries of the spectrum. Here, we give two examples of states progressively rotated around the $\hat{z}$ axis on the Bloch sphere by varying $\phi$ for two fixed values of $\theta$. The states are marked by the orange and green stars in Fig~\ref{fig_sup:other_init}(a). Both sets of states show progressively increasing energies/effective temperatures. Their trace distance dynamics, obtained by time evolution under the mixed field Ising Hamiltonian of Eq.~(1) of the main text, is plotted in Fig~\ref{fig_sup:other_init}(b)-(c). This shows the QME as $\phi$ is increased, whereas the IPR of the states progressively decreases with $\phi$ (inset), consistent with the results for Type 1 and 2 states in the main text. Since the IPR is anti-correlated with the initial trace distance, we expect to see the QME for \textit{any} two choices of $\theta,\phi$ in this model, as long as we are away from the edges of the spectrum.

\begin{figure}[tbh]
    \includegraphics[width=1\textwidth]{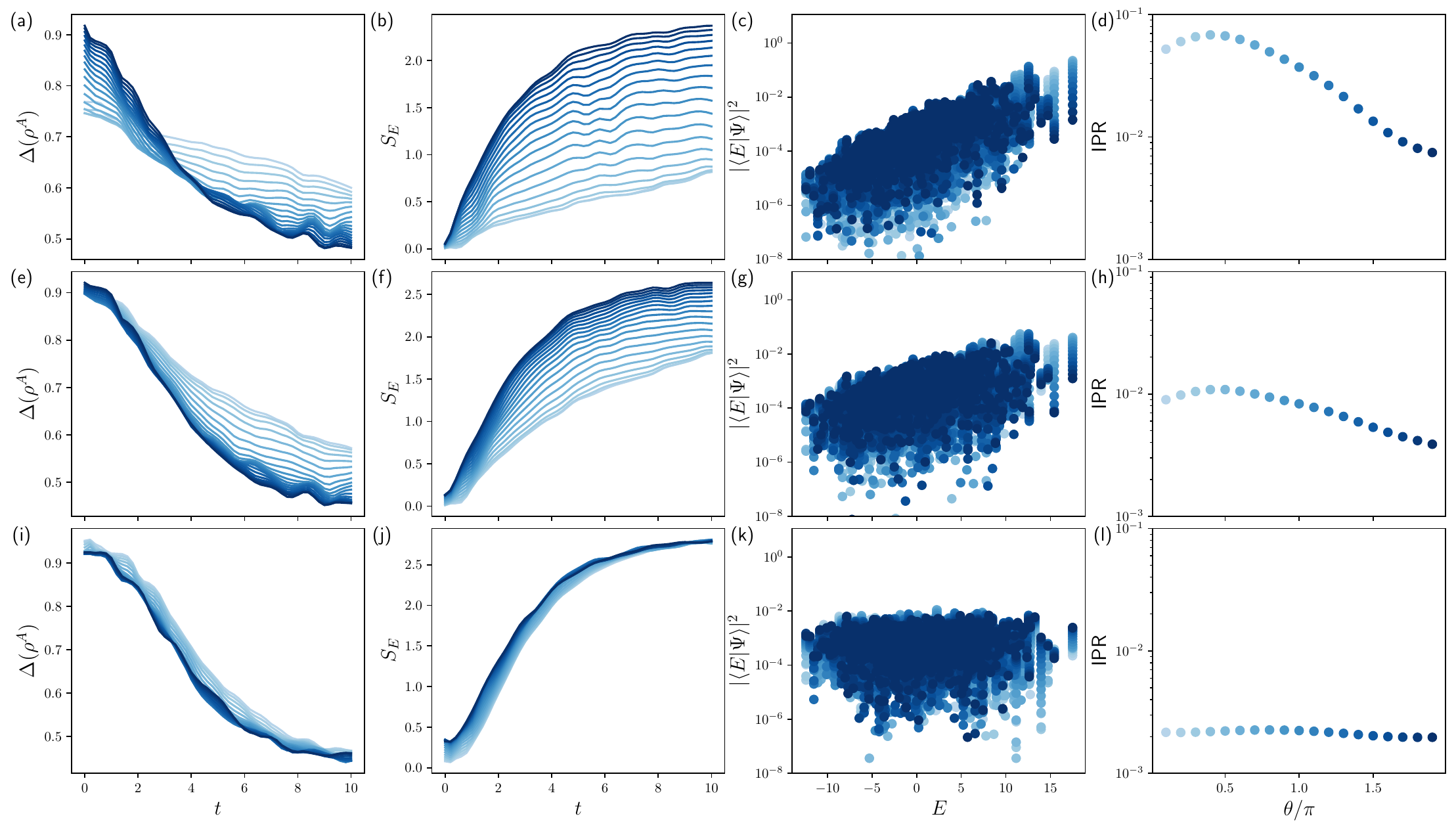}
    \caption{QME for weakly entangled initial states in the mixed field Ising model. Trace distance dynamics, entanglement entropy dynamics, overlaps with energy eigenstates and IPR of initial states which undergo random entangling evolution for $\delta t = 0.2$ (a)-(d), $\delta t = 0.3$ (e)-(h) and $\delta t$ = 0.5 (i)-(l). All data is for system size $N =10$ obtained by exact diagonalization.}
    \label{fig_sup:weak_entanglement}
\end{figure} 

For the time-independent Hamiltonians considered in the main text, we have focused on product states to demonstrate the QME, however, weakly-entangled initial states were required to observe the QME in the Floquet model. This raises the question: is the QME for time-independent Hamiltonians also sensitive to the entanglement in the initial state? Moreover, does the IPR remain a faithful criterion for weakly-entangled initial states? Here, we answer this  question by looking at the dynamics of the mixed field Ising model in Eq.~(1) of the main text for weakly entangled initial states.
We consider ferromagnetic states tilted around the $y$-axis, as defined in Eq~(\ref{suppeq:tiltedFM}) for $\alpha = y$. We entangle these initial states through the action of a local unitary operator of the form $e^{-i \delta t \hat{H}_R}$, where $\hat{H}_R = \sum_i \hat{h}_i$, and each $\hat{h}_i$ is a Haar-random operator defined on two sites. After a random evolution up to time $\delta t$, we evolve the resulting states with the mixed field Ising Hamiltonian in Eq.~(1) of the main text. The resulting dynamics of the subsystem trace distance, entanglement entropy, overlaps with energy eigenstates and the IPR are shown in Fig.~\ref{fig_sup:weak_entanglement} for $\delta t = 0.2,0.3, 0.5$. For a small $\delta t$, the initial states are weakly entangled and the QME is visible with a large variation in the dynamics and energy distributions. With increasing $\delta t$, the initial states are ``homogenized'' by the random Hamiltonian leading to almost identical dynamics, thus the QME is absent. Thus, the IPR correlates with the QME for weakly-entangled initial states, even for states that break translation invariance.  

\section{QME with a fixed initial state}

The conventional setup of the QME consists of a fixed Hamiltonian and varying initial states. Here, we consider an alternative setup with a fixed initial state that is quenched using different Hamiltonians, leading to the QME. We exemplify this using the random Hamiltonian in Eq.~(7) of the main text, where the three components $\mathbf{v}_i = (v_x,v_y,v_z)_i$ are real numbers drawn uniformly from the window $[-W,W]$ and $J_\mathrm{H}$ is kept uniform and controls the strength of the Heisenberg term.  We fix the initial state to be the ferromagnetic state $\ket{\mathrm{FM}}\equiv \ket{\theta=0,\phi}$ in  Eq.~(1) of the main text. Any local term of the form $(\boldsymbol{\sigma}_i \times \boldsymbol{\sigma}_{i+1})\cdot\mathbf{\hat v}$ has a zero expectation value for $\ket{\mathrm{FM}}$, making $\ket{\mathrm{FM}}$ an infinite temperature state for such Hamiltonians. We tune the energy density of $\ket{\mathrm{FM}}$ via the strength of $J_\mathrm{H}$ of the Heisenberg term. 

\begin{figure}[tb]
    \includegraphics[width=0.7\textwidth]{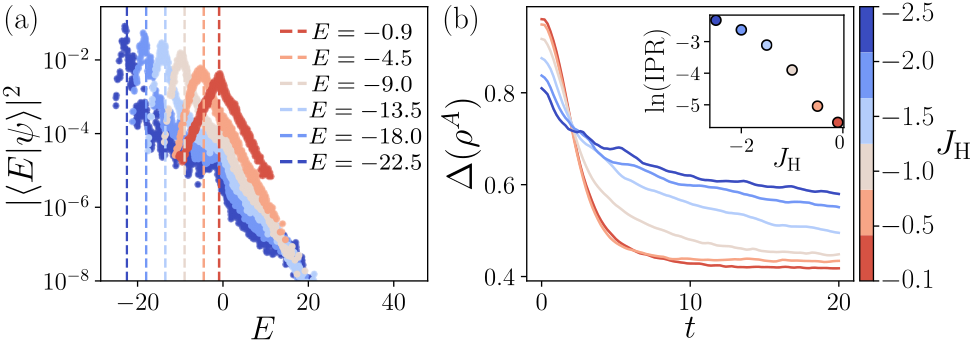}
    \caption{QME in the random model, Eq.~(7) of the main text, with a fixed initial state $\ket{\mathrm{FM}}\equiv \ket{\theta=0,\phi}$. (a) Averaged overlaps of $\ket{\mathrm{FM}}$ state with energy eigenstates as a function of their energy. Different colors represent different values of $J_\mathrm{H}$ [denoted in the legend of panel (b)], which translate into the average values of the energy quoted  in the legend. (b) Trace distance dynamics for the $\ket{\mathrm{FM}}$ state and different values of  $J_\mathrm{H}$, with the corresponding IPR values given in the inset. The crossing associated with the QME is visible as the effective temperature of the state is tuned by deforming the Hamiltonian. The inset shows the IPR for the same states as a function of $f$. All data is for the system size $N=10$ and fixed disorder window $W = 1$. 
    }
    \label{fig:random_hamiltonian_overlaps_SM}
\end{figure}
In Fig.~\ref{fig:random_hamiltonian_overlaps_SM}(a), we show the overlaps of $\ket{\mathrm{FM}}$ with energy eigenstates of $H_\mathrm{R}$, for a few choices of $J_H$ and each averaged over 50 random realizations. For large negative $J_\mathrm{H}$, the overlap peaks near the ground state, and in the limit $J_\mathrm{H} \to -\infty$, $\ket{\mathrm{FM}}$ becomes the ground state of ${H}_\mathrm{R}$. As $J_\mathrm{H}$ increases toward zero, the overlap peak shifts to the middle of the spectrum, moving the state toward infinite temperature. 
This coincides with a flattening of the distribution, as reflected in the decreasing \text{IPR} shown in the inset of Fig.~\ref{fig:random_hamiltonian_overlaps_SM}(b), marking delocalization over a larger number of eigenstates. Indeed, QME is evident in the trace distance dynamics in Fig.~\ref{fig:random_hamiltonian_overlaps_SM}(b), where increasing $J_\mathrm{H}$ leads simultaneously to higher initial $\Delta(\rho^A)$ values and faster decay rates, resulting in a crossing. 

\section{Strong vs. weak thermalization as a form of the QME}

Differences in the rate of thermalization of product states in non-integrable models, dubbed ``strong'' and ``weak'' thermalization, were noted in Ref.~\cite{Banuls2011} predating recent QME developments. Here, we argue that the results of Ref.~\cite{Banuls2011} can be interpreted as a manifestation of the QME dynamics, which can be explained by the simple criteria based on energy distributions introduced in the main text. 

\begin{figure}[tbh]
    \includegraphics[width=\textwidth]{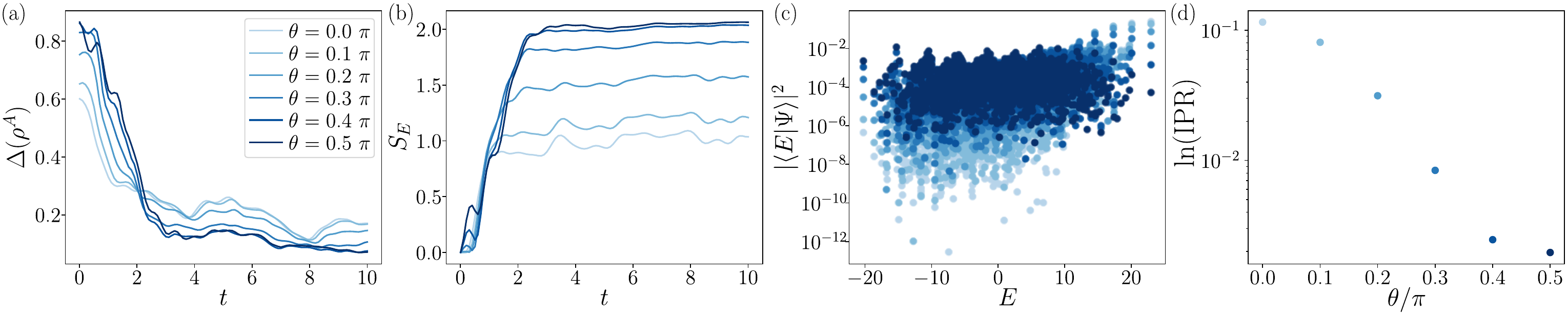}
    \caption{Strong vs. weak thermalization in the mixed field Ising model~\cite{Banuls2011} as a  manifestation of  the QME. We consider ferromagnetic states in Eq.~(\ref{suppeq:tiltedFM}) rotated about the $x$-axis. 
     (a) Trace distance dynamics. (b) Entanglement entropy dynamics. (c) Overlaps with energy eigenstates. (d) IPR as a function of tilt angle $\theta$. Data is for system size $N = 10$ obtained by exact diagonalization.
    }
    \label{fig_sup:strong_weak_thermal}
\end{figure}

We consider the same mixed field Ising model as in the main text 
and set the field strengths to $(J_{zz}, h_x,h_z) = (1, -1.5,0.5)$, as done in Ref.~\cite{Banuls2011} (see also Ref.~\cite{maceira2024thermalizationdynamicsclosedquantum}). Excluding the additional boundary terms, the Hamiltonian has a reflection symmetry about the center of the chain, which we resolve by restricting the Hilbert space to states with parity $+1$.
As an example of a strongly-thermalizing state, Ref.~\cite{Banuls2011} considered the product state of spins polarized along the $y$-axis, which is effectively at infinite temperature $\beta=0$. On the other hand, an example of a ``weakly-thermalizing'' state was the product state of spins pointing along the $z$-direction, whose effective temperature is $\beta \approx 0.7275$. In Fig.~\ref{fig_sup:strong_weak_thermal} we demonstrate that these two states are associated with the QME. We can tune between them by taking the ferromagnetic states in Eq.~(\ref{suppeq:tiltedFM}) and performing the rotation by an angle $\theta$ around the $\hat{x}$-axis, with the two states recovered in the limits $\theta=0$ and $\theta=\pi/2$. We fix the subsystem $A$ to be the central 3 sites of the chain and calculate the trace distance dynamics, entanglement entropy, overlaps with energy eigenstates and the IPR in Fig.~\ref{fig_sup:strong_weak_thermal}. Upon increasing $\theta$, QME dynamics is evident the initial trace distance and its rate of decay increases simultaneously, accompanied by an increasing growth rate and saturation value of entanglement entropy. We observe that the IPR of these states also steadily decreases with increasing $\theta$, consistent with other results in the main text.

\end{document}